\documentclass[preprint,5p,number,times,sort&compress]{elsarticle}

\usepackage{lineno}
\usepackage[colorlinks,
            linkcolor=blue, 
            anchorcolor=blue,
            citecolor=blue, 
            ]{hyperref}

\usepackage{color}
\usepackage{amsmath}
\modulolinenumbers[1]

\usepackage[utf8]{inputenc}
\usepackage{graphicx}
\usepackage{amsmath}
\usepackage{booktabs}
\usepackage{subfigure}

\usepackage{amssymb}
\usepackage{ulem}
\usepackage{multirow}
\usepackage{soul}
\setstcolor{red}

\usepackage{arydshln}

\journal{EPJC}

\begin{document}

\begin{frontmatter}

\title{Machine-Learning based photon counting for PMT waveforms and its application to the improvement of the energy resolution in large liquid scintillator detectors}

\author[firstaddress,thirdaddress]{Wei Jiang\texorpdfstring{{}}}
\cortext[mail]{Corresponding author}
\author[secondaddress]{\texorpdfstring{{}Guihong Huang \corref{mail}}{}}
\ead{huanggh@wyu.edu.cn}
\author[thirdaddress]{\texorpdfstring{{}Zhen Liu\corref{mail}}{}}
\ead{liuzhen@ihep.ac.cn}
\author[thirdaddress]{\texorpdfstring{{}Wuming Luo\corref{mail}}{}}
\ead{luowm@ihep.ac.cn}
\author[thirdaddress]{Liangjian Wen}
\author[secondaddress]{Jianyi Luo}

\address[firstaddress]{School of Physical Sciences, University of Chinese Academy of Science, Beijing 100049, China}
\address[secondaddress]{Wuyi University, Jiangmen 529020, China}
\address[thirdaddress]{Institute of High Energy Physics, Chinese Academy of Sciences, Beijing 100049, China}

\begin{abstract}
Photomultiplier tubes (PMTs) are widely used in particle experiments for photon detection. PMT waveform analysis is crucial for high-precision measurements of the position and energy of incident particles in liquid scintillator (LS) detectors. 
A key factor contributing to the energy resolution in large liquid scintillator detectors with PMTs is the charge smearing of PMTs.
This paper presents a machine-learning-based photon counting method for PMT waveforms and its application to the energy reconstruction,
using the JUNO experiment as an example.
The results indicate that leveraging the photon counting information from the machine learning model can partially mitigate the impact of PMT charge smearing and lead to a relative 2.0\% to 2.8\% improvement on the energy resolution in the energy range of [1, 9]~MeV.

\end{abstract}


\end{frontmatter}

\section{Introduction}

Liquid scintillators (LS) and photomultiplier tubes (PMT) play important roles in neutrino experiments, as evidenced in projects like KamLAND~\cite{KamLAND, KamLAND:2008dgz}, Borexino~\cite{Borexino, Borexino:CNO}, Daya Bay~\cite{Cao:2016vwh, DayaBay:2012fng, DayaBay:2022orm, DayaBay:2022jnn}, RENO~\cite{RENO,RENO:2012mkc} and Double Chooz~\cite{DoubleChooz,DoubleChooz:2011ymz}.
These LS neutrino detectors typically offer low energy thresholds of sub-MeV and good energy resolutions, which are ideal for studying neutrinos originating from sources including nuclear reactors, supernova, the Earth or the sun.
The next generation LS detector, Jiangmen Underground Neutrino Observatory (JUNO)~\cite{juno, junodet,juno:ppnp}, is equipped with 20 kton of LS, 17,612 20-inch PMTs and 25,600 3-inch PMTs within the central detector. 
Its primary goal is to determine the neutrino mass ordering (NMO), by precisely measuring the energy spectrum of reactor antineutrinos with a target energy resolution of 3\% @ 1~MeV. 
Reactor antineutrinos are detected in LS detectors via the Inverse Beta Decay (IBD) process, in which a positron and a neutron are produced in the final state. The positron will usually deposit its kinetic energy quickly and then annihilate with an electron, yielding the prompt signal. 
Meanwhile the neutron will dissipate its energy and eventually be captured by either hydrogen or carbon nuclei, yielding the delayed signal roughly 200~$\mu$s later. Taking advantage of the correlated prompt-delay signals, IBD events can be selected efficiently. The visible energy of the positron needs to be reconstructed in order to deduce the energy of the incoming antineutrino. Many factors will contribute to the energy resolution of positrons in JUNO, including the light yield of the LS, Cherenkov photons, non-uniform energy response of the detector, photon contamination from the PMT dark noise and more. A comprehensive breakdown of the energy resolution in JUNO can be found in Ref.~\cite{JUNORes}.

As indicated by Fig.~8 from Ref.~\cite{JUNONMO}, the energy resolution is crucial for the NMO sensitivity in JUNO, a relative 3.3\% improvement on the targeted 3\%@1~MeV energy resolution would roughly result in a 11.4\% reduction in the data taking time to reach 3$\sigma$ significance. To achieve the unprecedented energy resolution in JUNO, the LS recipe has been studied thoroughly~\cite{junoLS} to optimize the light yield. Meanwhile PMTs with world-leading Quantum Efficiency~\cite{junoLPMT} have been developed to further increase the number of detected photons. Prior studies have extensively explored vertex~\cite{Liu,Li} and energy reconstructions~\cite{Wu,Huang}. 
A data-driven simultaneous vertex and energy reconstruction method 
that combines the charge and time information of PMTs has been developed. 
It is used to predict the latest energy resolution~\cite{JUNORes}, which is subsequently used in the NMO analysis in JUNO~\cite{JUNONMO}. 
Comparing to previous studies, this method~\cite{QTMLE} provides a more realistic and accurate expected PMT response of positron events in JUNO. 
Given the importance of the energy resolution, 
an analysis was conducted to decompose the energy resolution and assess the impact of each contributing factor as shown in Fig.~15 from Ref.~\cite{JUNORes}, identifying the charge resolution of a single Photo-Electron (PE) in PMTs as the dominant contributor.
 
Large-area PMTs are often used in neutrino or dark matter experiments to detect photons~\cite{superK, IceCube}.
Due to the stochastic property of the amplification process, these PMTs have an intrinsic charge resolution of about 30\%~\cite{junoLPMT}. 
This causes an uncertainty on the charge-based number of PEs (nPEs) estimation and eventually on the energy resolution,
which is a common issue for low-energy experiments using large-area PMTs.
Meanwhile, Machine Learning (ML) has been more and more widely used in neutrino experiments such as MicroBooNE~\cite{MB1, MB2}, JUNO~\cite{JUNO_ML_atm, JUNO_ML_vertex, mlVertex2, JUNO_ML_E}, Super-K~\cite{SKML}, IceCube~\cite{IC1}, PandaX-4T~\cite{PandaX}, XENONnT~\cite{XENON}, EXO-200~\cite{EXO1, EXO2, EXO3}, in the areas of event reconstruction, particle identification or event selection. While these studies have demonstrated the great potential of ML on enhancing the detector performance at the event level, the application of ML to detector sensor units such as PMTs has barely been attempted. 
In this paper, we propose a ML approach to directly predict the nPEs from PMT waveforms  while reduce the smearing effect of the PMT charge. This machine-learning-based PE counting method demonstrates high accuracy, particularly when the number of detected PEs is low. When this method is applied to the energy reconstruction in low-energy PMT-based experiments such as JUNO, it partially mitigates the impact of PMT charge resolution on energy reconstruction, and thus leads to an improved energy resolution.

In the rest of the paper, Sec.~\ref{sec:MC} introduces the Monte Carlo samples used in this study. Sec.~\ref{sec:MLPE} presents the details of the machine-learning based photon counting method for PMT waveforms and its performances. Sec.~\ref{sec:Erec} describes how the photon counting information can be applied to the energy reconstruction and used to improve the energy resolution in JUNO. The discussion of an alternative method and conclusion are presented in Sec.~\ref{sec:discussion} and Sec.~\ref{sec:summary}, respectively.

\section{Monte Carlo Samples}
\label{sec:MC}
 The data used for evaluating the performance and efficacy of the reconstruction methods in this paper were generated by the official Monte Carlo simulation software of JUNO~\cite{Sim}. The data provides a comprehensive representation of the main components of the detector, including the properties of the photosensitive devices and the detection medium. The simulation accurately captures the response characteristics of the PMTs and the LS, providing a detailed depiction of the PMT waveforms, pulse shapes, time smear, charge smear, and the electronic noise.

\subsection{Detector}

The central detector (CD) of JUNO utilizes 20 kton of LS, contained in a 35.4-meter diameter, 12-centimeter thick acrylic sphere. 
The JUNO LS is composed of purified linear alkylbenzene (LAB) as the solvent, 2.5 g/L 2,5-diphenyloxazole (PPO) as the fluor, and 3 mg/L bis-MSB as the wavelength shifter~\cite{juno:ppnp}. This particular formulation is tailored for the optimal performance in JUNO~\cite{junoLS}, exhibiting high transparency and light yield. It achieves an attenuation length exceeding 20 m~\cite{juno:ppnp}, 
with the observed light yield of about 1665 p.e./MeV, where p.e. represents the unit of nPEs. The fast and slow components of the LS have a luminescence decay time of $\sim$5~ns and $\sim$20~ns, respectively~\cite{LStime}. 
Meanwhile, various photon production and propagation processes including scintillation, quenching, Cherenkov radiation, absorption and re-emission. Rayleigh scattering, reflection and refraction between medium boundaries, are also implemented in the simulation, with key parameters obtained from table top experiments. More details can be found in Ref.~\cite{JUNORes}.

The photo-sensors of CD consist of 5,000 20-inch Dynode-PMTs, 12,612 20-inch MCP (Microchannel Plates)-PMTs~\cite{Wen:2019sik} and 25,600 3-inch PMTs \cite{junoSPMT}. The testing results of the running performances of the 20-inch PMTs are reported in Ref.~\cite{junoLPMT}. Dynode-PMTs exhibit better time and charge resolution,
while MCP-PMTs offer better detection efficiency. 
Collectively, they deliver complementary performance. 

Besides the LS and PMTs~\cite{LPMT1,LPMT2}, the details of the other components of the CD can be found in Ref.~\cite{juno:ppnp}. 

\subsection{PMT waveforms}
When a particle deposits energy in LS, photons are emitted. Part of these photons then propagate through LS and are detected by the PMTs, while others are absorbed by LS. 
Both the number and hit times of these photons carry information about the vertex and energy of the particle. Therefore, accurately extracting these observables from the PMT signals is crucial.

\begin{figure}[htbp]
    \centering
    \includegraphics[width=0.45\textwidth]{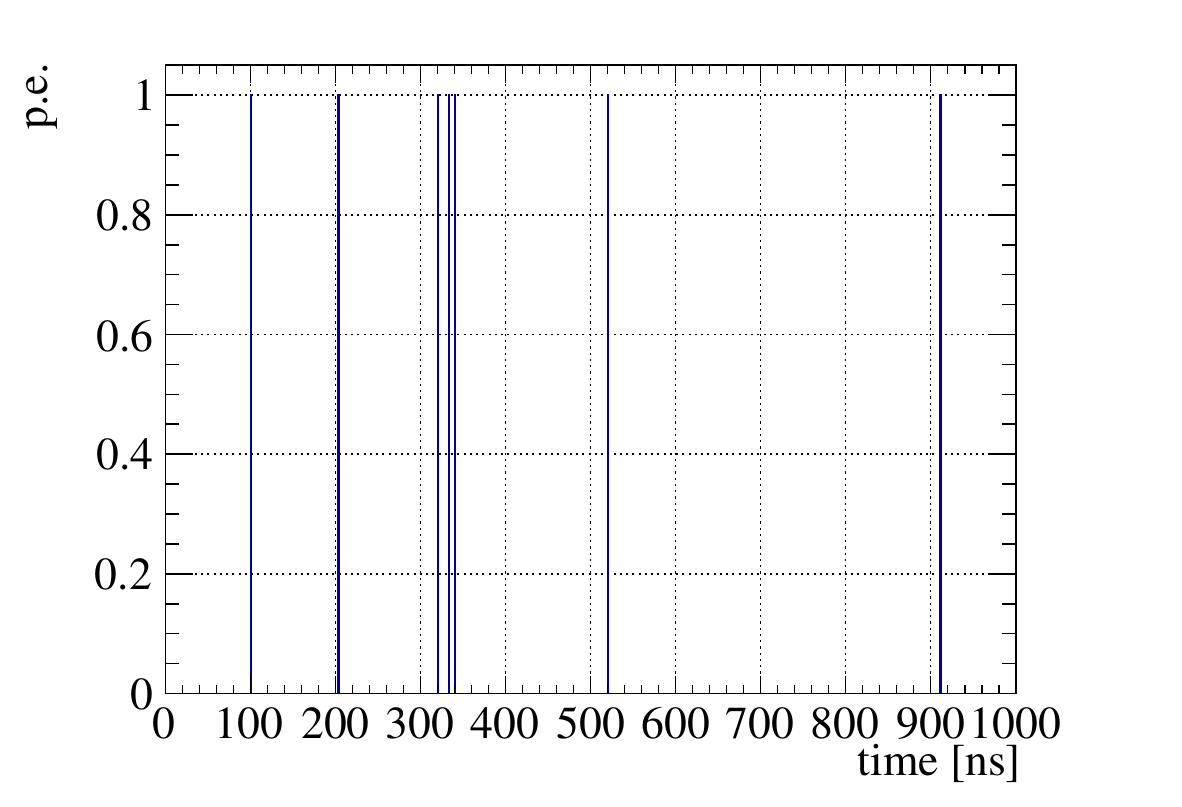}
    \includegraphics[width=0.45\textwidth]{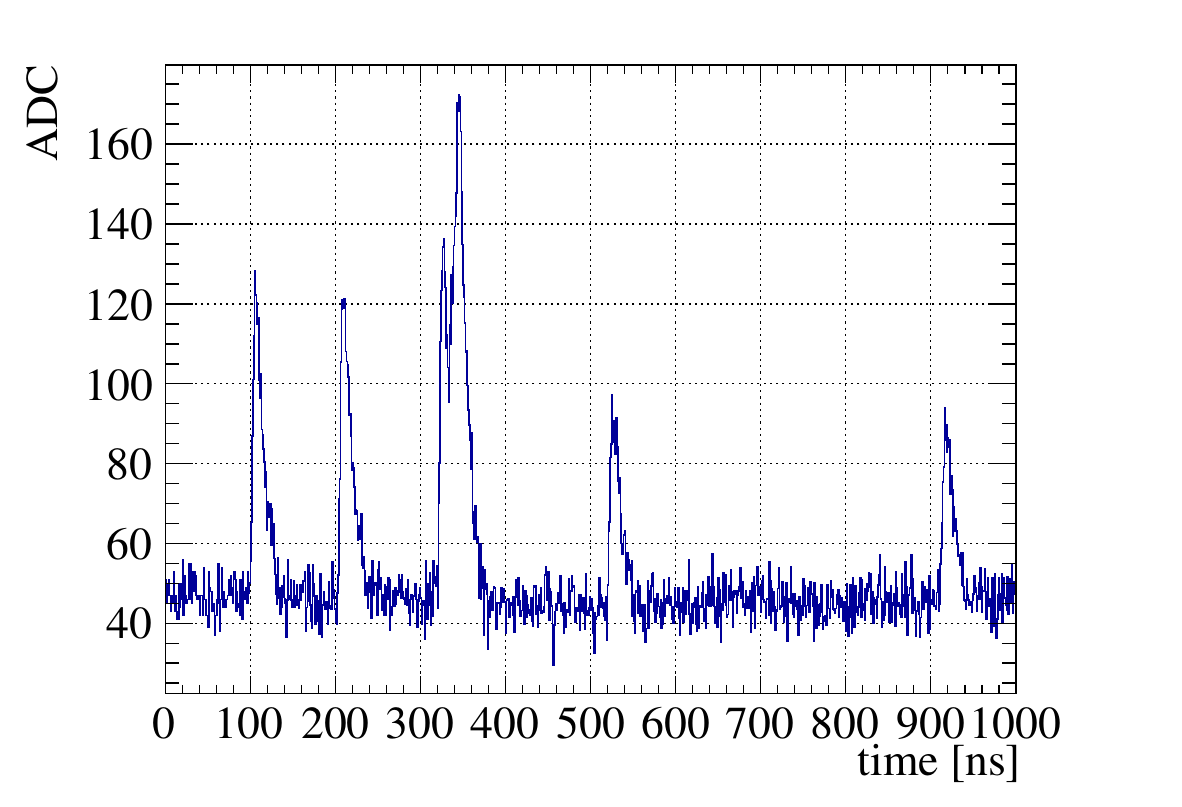}
    \includegraphics[width=0.45\textwidth]{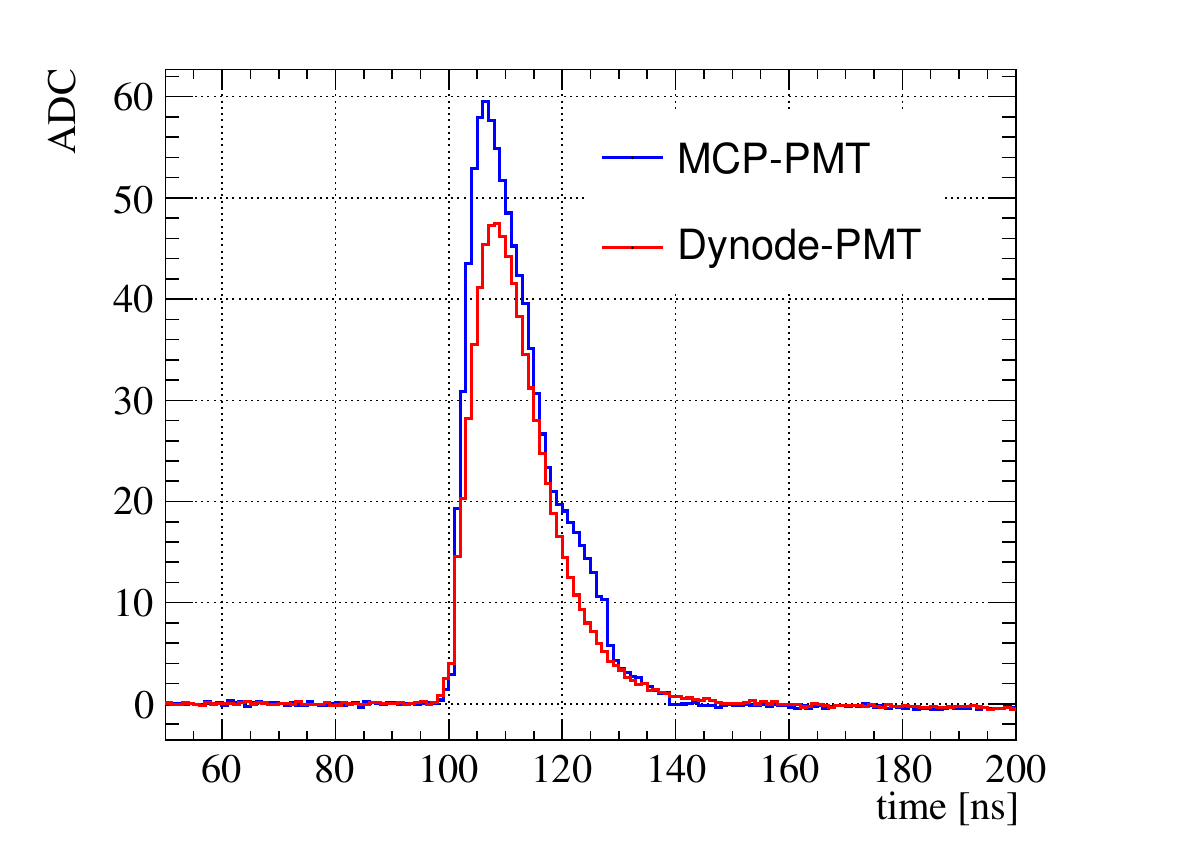}
    \caption{An example of an ideal waveform (top) containing 7 p.e. and the corresponding raw waveform (middle). The ideal waveform clearly exhibits the hit time of each individual PE. Due to the stochastic nature of the photoelectron amplification process, the hit time of the PEs in the raw waveform has become blurred, leading to the phenomenon of pile up among adjacent PEs. The typical PMT response for a single PE is shown in the bottom plot, for both types of PMTs.}
    \label{fig:raw_WF_7PE}
\end{figure}

Once a photon reaches a PMT, it may knock out a photoelectron in the PMT photocathode.
The photoelectron then undergoes amplification through either dynodes or microchannel plates. 
The resulted current signal at the PMT anode is converted into a voltage signal through electronic modules, and eventually recorded as a PMT waveform signal by a custom Analog to Digital Converter (ADC) operating at 1~GHz frequency~\cite{junoElec, FADC}. 
An ideal waveform, illustrated in the top plot of Fig.\ref{fig:raw_WF_7PE}, clearly shows the hit times of each photon, 
allowing for the accurate determination of nPEs as 7.
However, in reality, the stochastic nature of the photoelectron amplification process primarily causes smearing of the hit time and charge of each PE. Additionally, the electronic modules will also produce a pulse response for each PE.
Taking all the effects into account, realistic waveforms, as exemplified in the bottom plot of Fig.\ref{fig:raw_WF_7PE}, obscure the accurate identification of 7 p.e. and their corresponding hit time.

As mentioned previously, as the primary records of the detector’s response to each event, the PMTs waveforms are of great importance for event reconstruction. 
The characteristics of the waveforms are mainly driven by the LS and PMT properties. In the following sections, a few key features will be discussed. 
\subsubsection{Pulse shape}

The shape of the PMT single photoelectron (SPE) response depends on the structure and operating conditions of the PMT, as well as the readout electronics. 
It typically features a main peak, overshoot, and reflections, with the full width half maximum (FWHM) of the pulse being about tens of nanoseconds~\cite{FADC}. 
The average SPE waveform of MCP-PMT is slightly wider compared to that of Dynode-PMT. 
In real data, it could be slightly different for each PMT. In the simulation data used in this paper, the average SPE waveform for each type of PMT, derived from batch testing results~\cite{junoLPMT}, is used to represent that PMT type for simplicity, as shown in the bottom plot in Fig.~\ref{fig:raw_WF_7PE}.

\subsubsection{Time smear}
The time interval between a photon hitting a PMT and the subsequent electrical signal output is a random variable mainly due to
the photoelectron amplification process.
This interval approximately follows a Gaussian distribution, characterized by the mean transit time (TT) and the standard deviation known as the transit time spread (TTS)~\cite{TTS}. Both TT and TTS parameters have been measured during the PMT batch testing.
It is also important to note that both TT and TTS are influenced by the photon's incident position on the PMT photocathode, which can be quantified by the zenith angle of the incident point. This dependency varies between Dynode-PMTs and MCP-PMTs due to their different structures and amplification principles. 
The measurements of TT and TTS, accounting for these dependencies, have been incorporated into the simulation.

\subsubsection{Charge smear}
The stochastic nature of the PMT photoeletron amplification makes the charge of each PE a random variable.
By collecting a large number of SPE waveforms, the SPE charge spectrum of each PMT can be obtained. This is typically achieved using a low-intensity light source or dark noise. 
The typical SPE charge spetra of Dynode-PMTs is close to a Gaussian distribution, while for MCP-PMTs it has a long tail due to second photoeletron emission of the micro channel plates~\cite{MCPPMT}.
The SPE charge spectra for each PMT based on the batch testing results have been implemented in the simulation.

\subsubsection{Electronic noise}
In the absence of photoelectrons or dark noise, the waveform readout from the PMT is characterized as electronic white noise. 
The amplitude of the white noise approximately follows Gaussian distribution, with the $\sigma$ approximately one-tenth that of a SPE signal in JUNO. 
In the simulation, the white noise is randomly generated using parameters of the Gaussian distribution extracted from PMT testing data.

\subsection{Data samples}
In practice, each PMT, even with the same type, exhibits slight differences. In the simulation, PMT parameters including dark noise rate, gain, SPE charge resolution, photon detection efficiency, TT and TTS, are derived from batch testing results and varying PMT by PMT, to closely approximate real data.
On the other hand, for each type of PMTs, the average SPE pulse shape derived from batch testing is used for simplicity.
The simulation events generated for training and evaluating the machine-learning-based photon counting method, as well as for assessing the reconstruction performance, are listed in Table~\ref{tab:SampleInfo}.
Eight sets of positron samples, each with different kinetic energies, are produced with events uniformly distributed in the CD. 
\begin{table}[!h]
\centering
\caption{List of the MC simulation samples.}
    \begin{tabular}{lccc}
\toprule
       Type     & \textbf{Kinetic Energy} & \textbf{Statistics} & \textbf{Position}\\
\midrule
     $e^+$ & \begin{tabular}[c]{@{}c@{}}$E_k$=(0, 0.5, 1, 2, \\ 3, 4, 5, 8) MeV\end{tabular} & 500k/set & uniform in CD \\
\bottomrule
\end{tabular}
\label{tab:SampleInfo}
\end{table}

\section{Machine-Learning-Based Photon Counting for PMT Waveforms }
\label{sec:MLPE}

Ideal PMT waveform reconstruction aims to discriminate and reconstruct the hit time of each PE. The precision of the nPE count is critical for energy resolution, whereas the accuracy of photoelectron timing indirectly affects energy resolution through its impact on vertex reconstruction.
Given the limitations in time and charge resolution, as well as the SPE pulse shape, reconstructing every PE accurately is extremely challenging, particularly when multiple PEs overlap with each other.
Various waveform reconstruction algorithms have been explored in the Daya Bay experiment~\cite{FADC}. The deconvolution method filters out high-frequency white noise and partially mitigates the effects of SPE pulse shape. Consequently, it achieves the best charge non-linearity correction among the tested algorithms, and has been used in previous
studies for JUNO~\cite{QTMLE, Li} to provide the crucial inputs for the event reconstruction, namely the charge and hit time of PMTs.
However, the intrinsic charge resolution for Dynode and MCP PMTs in JUNO is about 28\% and 33\% respectively.
It is found to have a large impact on
the energy resolution in JUNO, as will be shown in Sec.~\ref{sec:Eres} later.

Machine learning is increasingly used in high energy physics, often addressing problems where traditional methods encounter limitations~\cite{JUNO_ML_atm, JUNO_ML_vertex, mlVertex2, JUNO_ML_E} .
In the case of PMT waveform reconstruction, as mentioned previously, directly discriminating and counting each PE could circumvent issues related to PMT charge resolution,
thus improving energy resolution.
There are ongoing studies aiming to achieve this photon counting mode for large-area PMTs by further improving the PMT time resolution.
However, identifying each PE and reconstructing its hit time poses significant challenges, even with the help of machine learning. 
Firstly, the number of PEs in a given waveform is not predetermined, making it challenging for machine learning models to adapt to a variable output dimension. Secondly, the limited time resolution and SPE response of PMTs complicate the separation of overlapping PEs.
Therefore, instead of attempting to identify and reconstruct the hit time of every PE, which is a regression task with an uncertain number of outputs, we aim to count the nPEs in each waveform using machine learning. This approach is more practical and feasible, as it simplifies the problem to a classification task.
The rest of this section introduces the details of applying machine learning models to count the nPEs in each waveform. 

\subsection{Training and testing samples}

\begin{figure}
    \centering
    \includegraphics[width=0.45\textwidth]{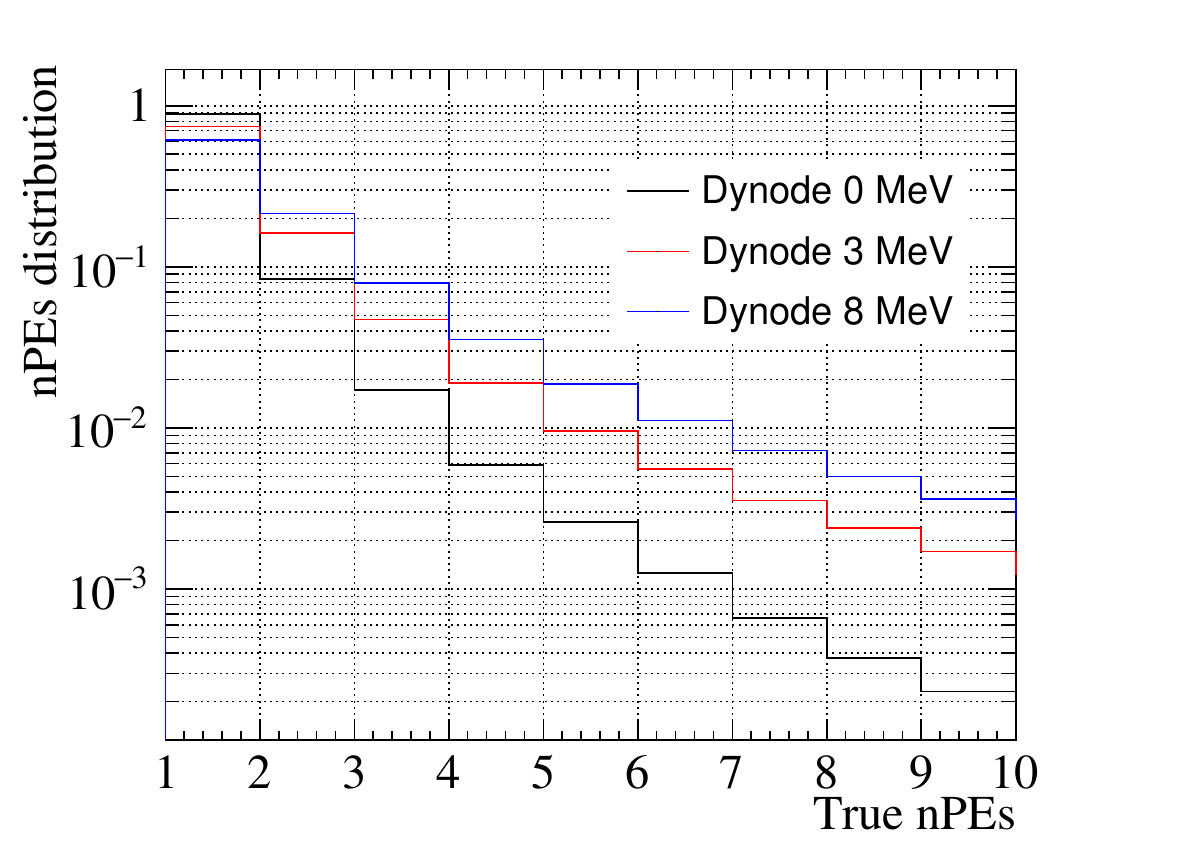}
    \caption{nPE distribution of Dynode-PMTs for positrons uniformly distributed in the CD. MCP-PMTs have similar distribution. For positron events with kinetic energy no more than 8 MeV, over 95\% of the PMTs will detect less than 5 p.e..}
    \label{fig:dyn_nPE1d}
\end{figure}

The training waveforms are randomly selected from positron events with different energies that are uniformly generated at different positions within the detector, as listed in Tab.~\ref{tab:SampleInfo}. 
To avoid any potential dependence on the positron vertex and energy,
each set of waveforms with $k$ ($k$ = 0, 1, ..., 8, $\geq$9) p.e. are evenly sampled with respect to both the energy $E$ and the cubic radius $r^3$ of the vertex.
Approximately 0.5 million waveforms are selected for each set to ensure a statistically robust training dataset.
The testing waveforms are selected in a similar way as the training waveforms, with about 1 million statistics for each set.
While the dynamic range for detected PEs of PMTs can exceed 9 , it is observed that waveforms with more than 9 p.e. constitute only about 1\% of the total at $E_k=8~$MeV, as shown in Fig.~\ref{fig:dyn_nPE1d}. 
Moreover, with an increase in nPEs, the probability of PE overlap increases, resulting in a rapid decline in the performance of the machine-learning-based PE counting, as will be demonstrated in subsequent sections.
Due to the different characteristics of the two types of PMTs, two separate waveform samples are prepared, one for each PMT type. 

\subsection{Machine-learning-based photon counting method}
The machine-learning-based photon counting method can be roughly broken down into three parts: the input, the machine learning network and the output.
For the input, a signal window of 420~ns is defined for the PMT waveforms, with further details provided in Sec.~\ref{sec:sw}.
The event is firstly reconstructed with the simultaneous vertex and energy reconstruction method in Ref.~\cite{QTMLE}. 
The distribution of residual hit times for all PMT hits is obtained by subtracting the calculated photon time of flight, which is derived from the reconstructed vertex and the PMT position. 
The starting time of the signal window is set at the rise time of the residual hit time distribution, and only waveform segments within the signal window are saved and used as input for the machine learning model.
Note that dark noise photons might occasionally fall into the 420~ns signal window, as described in Sec.~\ref{sec:sw}. However we do not differentiate the origins of the photons, and the contribution from dark noise is included in the true nPEs.

Photon hits on PMTs possess distinguishing characteristics in the PMT waveforms, such as pulse height, pulse width, charge, and time of arrival, allowing for the potential identification of the number of photons from each waveform. This task resembles challenges in industries such as speech recognition where sound waveforms are analyzed to distinguish individual speakers. 
In recent years, machine learning has significantly revolutionized speech recognition, achieving superior performance. Among the pioneering models in this field, RawNet~\cite{Rawnet} stands out as one of the state-of-the-art machine learning networks. Inspired by these advancements, we have chosen RawNet for photon counting in PMT waveforms. The architecture of our customized RawNet is shown in Tab.~\ref{tab:rawnet_struct}. With respect to the original model from Ref.~\cite{Rawnet}, the output shape is modified for some layers and (2+2) Residual blocks are used instead of (2+4).
In addtion, the hyperparameters of our model are summarized in Tab.~\ref{tab:rawnet_parameter}.

\begin{table}[t]
 \caption{Modified RawNet architecture. For convolutional layers, numbers inside parentheses refer to filter length, stride size, and number of filters. For gated recurrent unit (GRU) and fully-connected (FC) layers, numbers inside the parentheses indicate the number of nodes. BN stands for batch normalization.}
  \centering
  \label{tab:table1}
  \begin{tabular}{l c c}
  \toprule
  \textbf{Layer} & \textbf{Input} & \textbf{Output shape}\\
  \toprule
  \multirow{2}{*}{Strided} & Conv(3,3,128) & \multirow{3}{*}{(128, 140)}\\
  \multirow{2}{*}{-conv}& BN & \\
  & LeakyReLU & \\
  \midrule
  Res block & 
    $\left \{
      \begin{tabular}{c}
      Conv(3,1,128)\\
      BN \\
      LeakyReLU\\
      Conv(3,1,128)\\
      BN\\
      \hdashline
      LeakyReLU\\
      MaxPool(3)\\
      \end{tabular}
    \right \}$
    $\times$2
    
  & (128, 46)\\
  \midrule
  Res block & 
    
    $\left \{
      \begin{tabular}{c}
      Conv(3,1,256)\\
      BN \\
      LeakyReLU\\
      Conv(3,1,256)\\
      BN\\
      \hdashline
      LeakyReLU\\
      MaxPool(3)\\
      \end{tabular}
    \right \}$ 
    $\times$2
  & (256, 1)\\
  \midrule
  GRU & GRU(1024) & (1024,)\\
  \midrule
  Speaker & \multirow{2}{*}{FC(128)} & \multirow{2}{*}{(128,)}\\
  embedding & & \\
  \midrule
  Output & FC(10) & (10,)\\
  \bottomrule
  \end{tabular}
  \label{tab:rawnet_struct}
\end{table}

\begin{table}
\caption{Hyperparameters of the model. Some of the hyperparameters were optimized using a coarse-to-fine grid search via supervised training. We began with a broad search to identify promising regions, then gradually refined the search range until performance stabilized.}
\centering
\begin{tabular}{ll}
\hline
\textbf{Parameter} & \textbf{Value} \\
\hline
Optimizer & Adam \\
Optimizer Parameters & $\left\{\begin{array}{l}
           \text{Betas: } \beta_1 = 0.9, \beta_2 = 0.999\\
           \text{Weight decay: } 10^{-4}
           \end{array}\right.$ \\
Learning Rate Strategy & One-cycle policy \\
Learning Rate Parameters & $\left\{\begin{array}{l}
            \text{Division factor: } 25 \\ 
           \text{Maximum learning rate: } 10^{-3}\\
           \text{Final division factor: } 10^5 \\
           \end{array}\right.$ \\

Activation & ReLU \\
Loss Function & CrossEntropyLoss \\   
Training Epochs & 30 \\ 
Batch Size & 128\\
Gradient Clipping & 0.1 \\ 
\hline
\end{tabular}
\label{tab:rawnet_parameter}
\end{table}

As mentioned previously, photon counting for PMT waveforms can be treated as a classification problem. Thus, the model’s output for one waveform is a probability vector $\vec{p}$ = \{$p_k$\},
where $k$ = 0, 1, ..., 8, $\ge$9, $p_k$ denotes the corresponding probability of having $k$ photons in the 420~ns waveform, and $\sum{p_k} = 1$.

To optimize resource usage, waveforms from the same type of PMTs are grouped and trained together on a single neural network.
We conducted a comparison between separate and joint training for waveforms of MCP-PMTs and Dynode-PMTs. Due to the different characteristics of the two PMT types, separate training demonstrated better performance compared to joint training.
For each type of PMTs, the corresponding ML model was trained for 30 epochs. With the 5 million waveforms in total in the training datasets, the average training time for each epoch is about 15 minutes. With the aid of early stopping, the validation loss starts to converge around epoch 20, while the training loss continues to decrease slowly. In the end the model from epoch 22 or 24 with the minimal validation loss was chosen for MCP and Dynode PMTs respectively.

\subsection{Performance}
\label{sec:PC_perform}

\begin{figure}[h]
    \centering
    \includegraphics[width=0.38\textwidth]{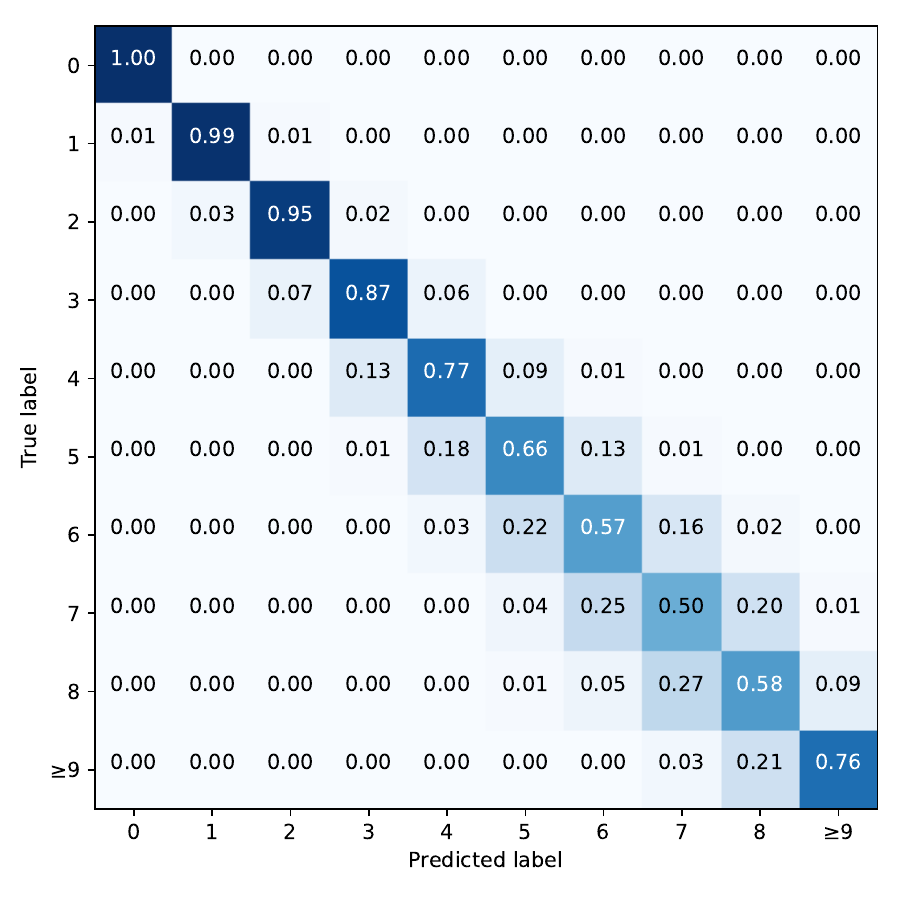}
    \includegraphics[width=0.38\textwidth]{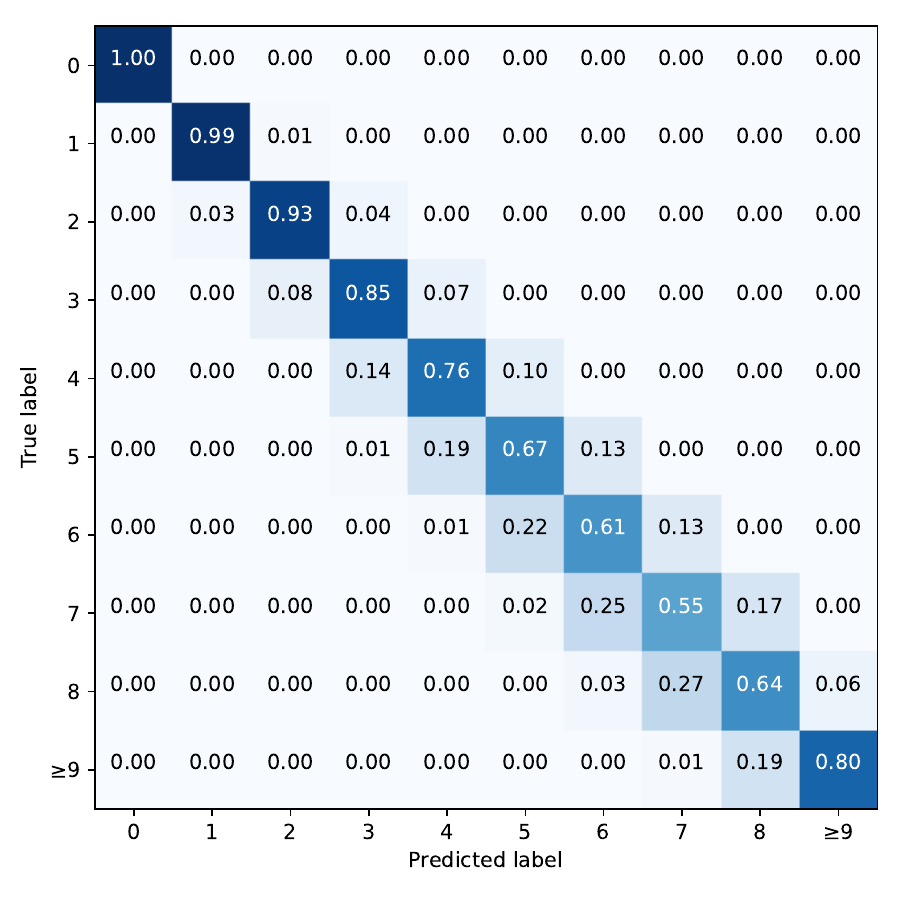}
    \caption{RawNet confusion matrix of MCP-PMT(upper) and Dynode-PMT(lower). The accuracy for 1 p.e. is about 99\% for both types of PMTs. It remains above 87\% for MCP-PMT and 85\% for Dynode-PMT for 3 p.e.. The counting accuracy decreases as the nPEs increases.} 
    \label{fig:trainCM}
\end{figure}

For every waveform, the RawNet model predicts the nPEs within the signal window as the category $k$ with the highest probability $p_{k}$. The accuracy of these predictions is evaluated by comparing them against the true nPE values, with the results summarized in a confusion matrix $C_{k k'}$. This matrix displays the likelihood of a waveform with $k$ p.e. being classified as $k'$ p.e., thus reflecting the model's accuracy. 
Separate confusion matrices are obtained for the two PMT types using the RawNet model, as shown in Fig.\ref{fig:trainCM}. 
In the low nPEs range, the machine-learning-based method for photon counting achieves high accuracy for both PMT types. In particular, the accuracy for 1 p.e. waveforms is around 99\%, and for 3 p.e. waveforms, it remains above 85\%.
As the number of PEs increases, the accuracy declines, which aligns with expectations: more PEs increase the probability of overlap in the waveform, making accurate prediction more difficult.
The reason for the unexpectedly high accuracy in the 8 p.e. and $\ge$ 9 p.e. waveforms is because the final category includes all waveforms with 9 or more p.e. This categorization leads to a concentration of more accurate results within that range, as well as a small probability for 8 p.e. waveforms to fall into that category. Additional checks were done in which the categorization extends to $\ge$ 19 p.e. instead of $\ge$ 9 p.e., then the accuracy for the 8 p.e. and 9 p.e. waveforms return to normal and follow the declining trend. In addition, various machine learning networks were tested and  similar performance were obtained.

\begin{figure}[h]
    \centering
    \includegraphics[width=0.38\textwidth]{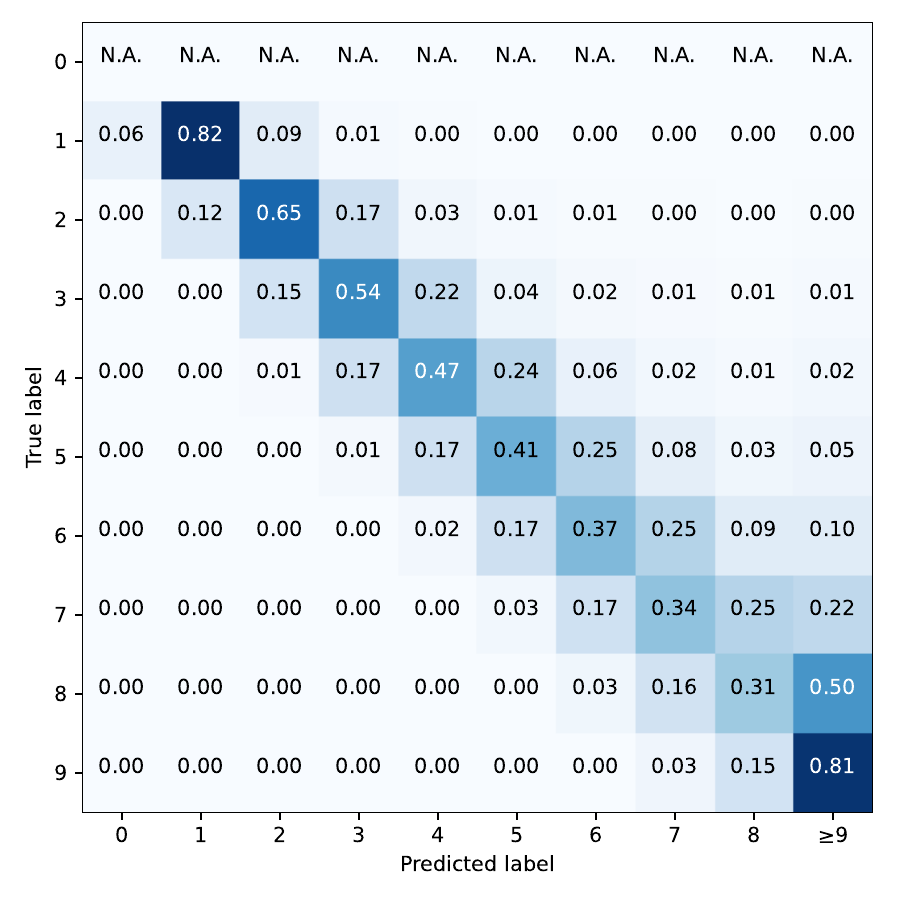}
    \includegraphics[width=0.38\textwidth]{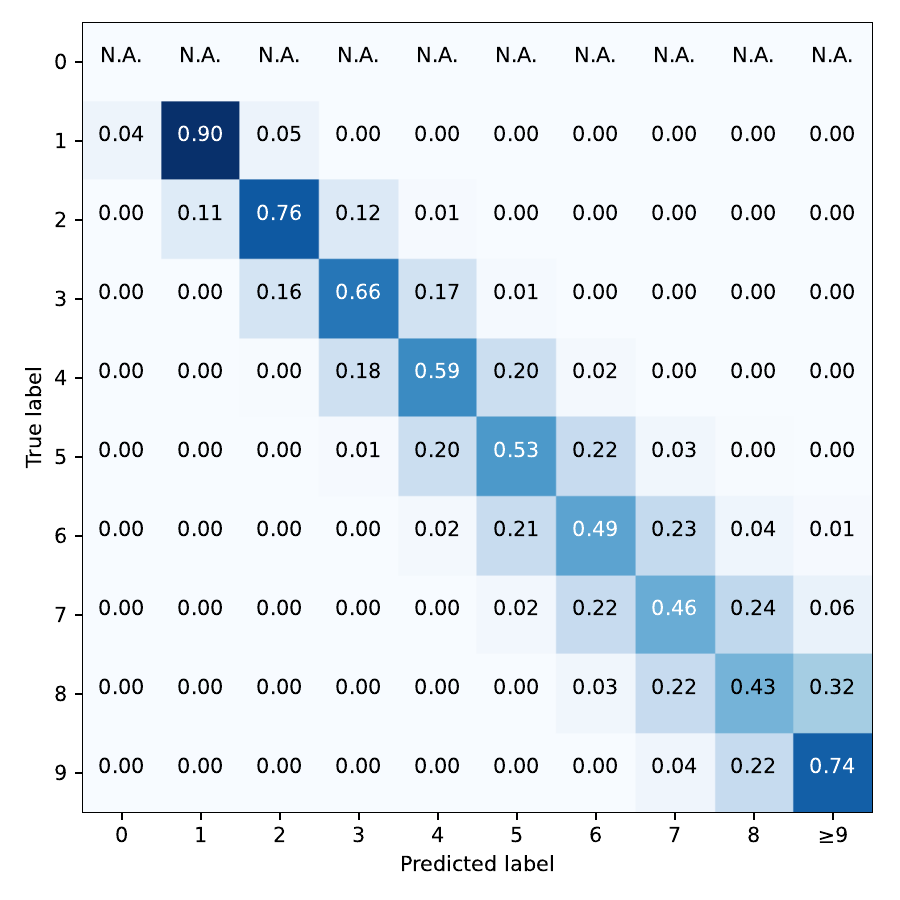}
    \caption{Confusion matrix based on charge classification of MCP-PMT(upper) and Dynode-PMT(lower). N.A. stands for not applicable. The counting accuracy is markedly inferior to that of RawNet. Besides, the counting accuracy of MCP-PMT is lower than that of Dynode-PMT, attributed to the fact that this method solely relies on charge information, while MCP-PMT has broader charge distribution.}
    \label{fig:deconvCM}
\end{figure}

In order to show that the ML based photon counting gives more accurate estimation of the nPEs than the charge obtained with applying the deconvolution method to the full waveform, the confusion matrix of a simple charge-based photon counting method is derived as shown in Fig.\ref{fig:deconvCM}. The charge spectrum for waveforms with k p.e. is utilized (k=0 case is not available). A PMT waveform is categorized into group $k'$ if the charge $q$ obtained with the deconvolution method meets the condition $k'-0.5 < q < k'+0.5 $. One can see that the accuracy of the charge-based photon counting method is noticeably inferior to that of RawNet.
Moreover, the asymmetry in the classification performance, in particular for the MCP-PMTs, is due to the non-Gaussian charge spectrum.

\begin{figure}[h]
    \centering
    \includegraphics[width=0.5\textwidth]{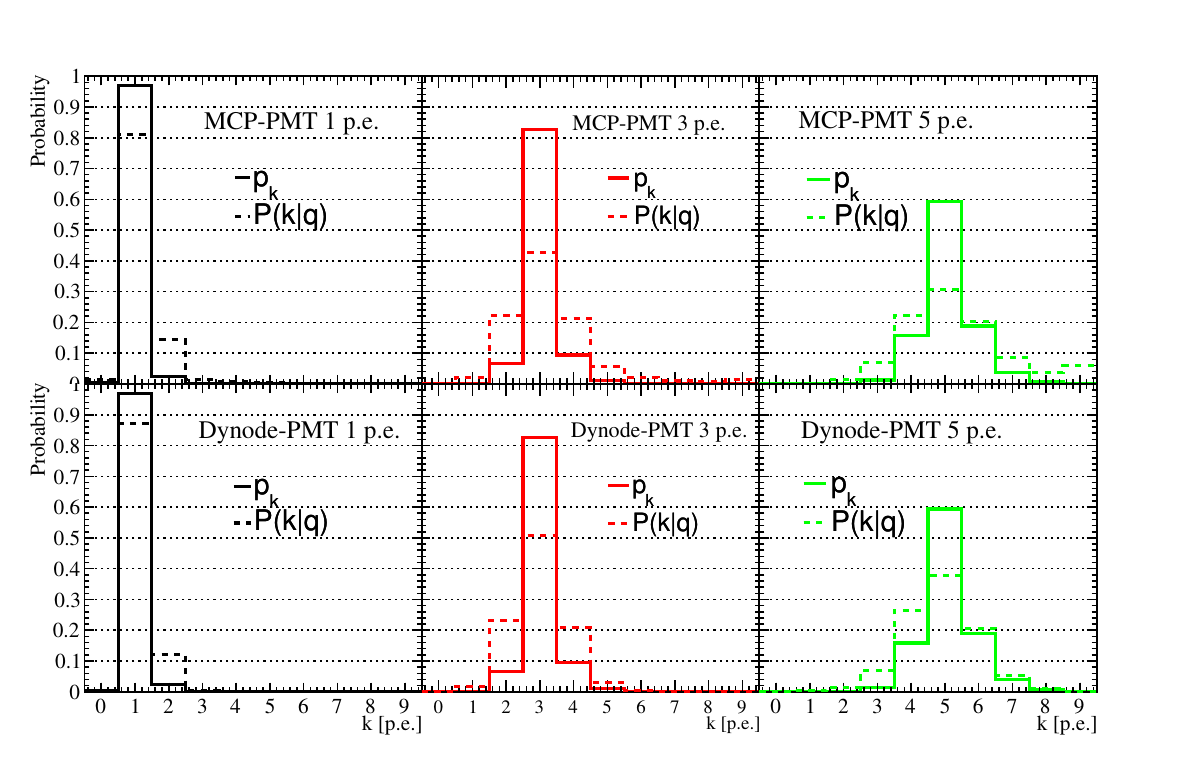}
    \caption{Comparison of the normalized \{$p_{k}$\} and \{$P(k|q)$\} distributions for waveform with nPEs = 1, 3 and 5 p.e.. Both distributions peak around n, but the former has a smaller spread. }
    \label{fig:pk_compare}
\end{figure}

Moreover, given the charge $q$ of a waveform, the probability of having k p.e. $P(k|q)$ can be calculated as $P(k|q)$ = 
$P_{Q}(q|k)$/$\sum_{k'} P_{Q}(q|k')$, where $P_{Q}(q|k)$ is the charge spectrum of k p.e. waveforms.
A direct comparison between \{$p_{k}$\} obtained from the ML method and \{$P(k|q)$\} from the charge inference above is shown in Fig.\ref{fig:pk_compare}. For each waveform with fixed n p.e., the corresponding \{$p_{k}$\} and \{$P(k|q)$\} values are filled into the solid and dashed histogram respectively. 
The normalized distributions for n=1, 3 and 5 are plotted as examples. One can see that the accuracy of \{$p_{k}$\} from RawNet is noticeably better than that of \{$P(k|q)$\} from the charge inference.

\section{Energy reconstruction with photon counting}
\label{sec:Erec}
Previous studies on energy reconstruction in JUNO have predominantly relied on PMT charge information alone~\cite{QTMLE}.
And we found that the charge smearing of PMTs is one of the key factors affecting energy resolution~\cite{JUNORes}.
However, as demonstrated in previous sections, the machine-learning-based photon counting method provides a more precise estimation of nPE, particularly in low PEs scenarios, than traditional charge-based estimation. This section introduces a novel approach that integrates both charge and nPEs information into the energy reconstruction. This method aims to further mitigate the impact of charge dispersion on energy resolution by leveraging photon-counting information obtained through machine learning techniques.

\subsection{Signal window}
\label{sec:sw}
For each triggered candidate signal event, a trigger window of 1250~ns is opened capturing the waveforms from all PMTs for subsequent event reconstruction. These waveforms contain PEs not only from the signal particle but also from the PMT dark noise.
With an average dark noise rate of 30~kHz across 17,612 PMTs, the estimated total dark noise PEs amount to approximately $30~\text{kHz} \times 1250~\text{ns} \times 17,612 = 660.45$ p.e., while a 1 MeV energy particle typically yields about 1665 p.e..
Consequently, the photon contamination from the PMT dark noise within the trigger window is one of the main contributing factors to the energy resolution. 
Given the random nature of PMT dark noise, in contrast to the temporally correlated photons from signal events, which predominantly occur within a few hundred nanoseconds, a signal window of 420~ns for each PMT is applied to the event reconstruction to reduce the impact of PMT dark noise. 
As mentioned previously, 
the start time of this signal window is based on the vertex reconstructed through the simultaneous vertex and energy reconstruction algorithm referenced in Ref.~\cite{QTMLE}.

\subsection{Usage of the photon counting information}
After applying the machine learning model to predict the nPEs within the 420~ns signal window for each PMT, a set of \{$p_k$\} is obtained, 
with $k$ ranging from 0 to $\ge$9. Here, $p_k$ represents the probability of predicting $k$ p.e., with the sum of all \{$p_k$\} equaling 1. 
Now, for each PMT, we have not only the charge and time information for the pulses within the signal window but also the photon counting information. The challenge lies in leveraging this photon counting data to enhance energy resolution.

Before tackling this challenge, it is insightful to examine the true nPEs distribution for the positron samples from Tab.~\ref{tab:SampleInfo} as shown in Fig.~\ref{fig:dyn_nPE1d}. As expected, with increasing positron energy, the nPEs distribution shifts rightward, reflecting higher photon emission. 
However, even at 8~MeV, 95.6\% of PMTs detect no more than 5 p.e..
Moreover, the confusion matrix for the ML-based photon counting model, shown in Fig.~\ref{fig:trainCM}, reveals that while the prediction accuracy for 1 or 2 p.e. is above 93\%, it decreases sharply with higher $k$ values. This decline aligns with expectations as the probability of pulse overlap increases with nPEs. 
Given that the intrinsic charge resolution of SPE is around 30\% and the width of SPE waveform spans approximately tens of ns, it becomes challenging to accurately predict the nPEs when there are overlapping pulses. 

We define the event reconstruction algorithm utilizing Charge and Time Maximum Likelihood Estimation (QTMLE), as described in Ref.~\cite{QTMLE}, as the reference case. For the 
$i$-th fired PMT in an event, the observable is the charge $q_i$ within the signal window. The corresponding likelihood function is constructed as follows:
 
\begin{equation}
\label{eq:likelihood_q}
    \mathcal{L}(q_i | \mu_i) = \sum_{k=1}^{+\infty} P_{Q}(q_{i}|k) P(k, \mu_i) 
\end{equation}

\noindent where $\mu_i$ is the expected nPEs for the i-th PMT, $P(k, \mu_i)$ is just the Poission probability of observing k p.e. given $\mu_i$ and $P_{Q}(q_{i}|k)$ is the charge pdf for k p.e..
In an ideal scenario, devoid of any charge smearing by the PMTs, the exact number of PEs 
$k_i$ within the signal window for the i-th PMT could be determined. Consequently, the likelihood function simplifies to:

\begin{equation}
\label{eq:likelihood_k}
    \mathcal{L}(k_i | \mu_i) =  P(k_i, \mu_i) 
\end{equation}

After applying the machine learning model to each PMT waveform of a positron event, 
we define $\kappa$ as the index for which $p_\kappa$ = Max( \{$p_k$\} ). Consequently, the 
i-th PMT is predicted to detect $\kappa_i$ p.e. with the highest probability.
There are different strategies for incorporating photon counting information into the reconstruction process.
Given the nPEs distribution and the high prediction accuracy of the machine learning model for small nPEs, one common principle for all the strategies would be to use the photon counting information, instead of the charge, for those PMTs with small $\kappa$ below a certain threshold $K_T$.

In the following part of this section, we will outline a strategy that employs the probability values \{$p_k$\} along with the corresponding nPEs values \{k\} on an individual PMT basis. For those PMTs satisfying the $\kappa \leq K_T $ requirement, all the \{$p_k$\} will be used to construct the likelihood function in a straightforward way, as illustrated below:

\begin{equation}
\label{eq:likelihood_p}
\begin{split}
   \mathcal{L}( \{p^i_{k}\} |\mu_i) =    
    \sum_{k=0}^{9} R_{K_{T}k} p^i_k P(k, \mu_i),
   R_{K_{T}k} = \sum^{K_T}_{\kappa = 0}{C_{k\kappa}}, \kappa_i \leq K_T,
\end{split}
\end{equation}

\noindent which is the superposition of the Poisson probability $P(k, \mu_i)$ weighted by the corresponding probability $p^i_k$. Note that for $p^i_9$, $P(9, \mu_i)$ is defined as $\sum_{k=9}^{+\infty} P(k, \mu_i)$.
The additional coefficient $R_{K_{T}k}$ is calculated as $\sum^{K_T}_{\kappa = 0}{C_{k\kappa}}$, which can be interpreted as the probability of a PMT with k p.e. satisfying the condition $\kappa \leq K_T$.
For PMTs with $\kappa > K_T $, a similar likelihood function to that in Eq.~\ref{eq:likelihood_q} will be employed, utilizing the PMT charge information. 
The complementary coefficient (1-$R_{K_{T}k}$)
represents the probability of a PMT with k p.e. satisfying the condition $\kappa > K_T$.

\begin{equation}
\label{eq:likelihood_qKT}
    \mathcal{L}(q_i | \mu_i) = \sum_{k=0}^{+\infty} (1 - R_{K_{T}k}) P_{Q}(q_{i}|k) P(k, \mu_i), \kappa_i > K_T. 
\end{equation}

By combining Eq.~\ref{eq:likelihood_p} and Eq.~\ref{eq:likelihood_qKT}, 
 we can construct the likelihood function for the i-th PMT with either the photon counting or the charge information, as follows:
 
 \begin{equation}
\label{eq:likelihood_mix}
\begin{split}
    \mathcal{L}(\{p^i_k\}, q_i | \mu_i) = 
    \sum^9_{k=0} R_{K_{T}k} p^i_k  P(k, \mu_i)  \theta(K_T- \kappa_i)~+ \\
    \sum_{k=0}^{\infty} (1-R_{K_{T}k})P_{Q}(q_{i}|k)  P(k, \mu_i)  \theta(\kappa_i - K_T-1)
\end{split}
\end{equation}

\noindent where the step functions $\theta(K_T-\kappa_i)$ and $\theta(\kappa_i-K_T-1)$ are the conditions that determine either charge $q_i$ or probabilities \{$p^i_k$\} will be used.
 The complete likelihood function for the event reconstruction can be constructed as follows:

\begin{equation}
\label{eq:EqQPTMLE}
\begin{split}
  &\mathcal{L}(\{p^i_{k}\};\{q_{i}\}; \{t_{i,r}\}|\mathbf{r},t_0,
    E_{\textrm{vis}}) = \\   
      &\prod_{i=1}^{17612} \left(  \sum^9_{k=0} R_{K_{T}k} p^i_k P(k, \mu_i)\theta(K_T- \kappa_i)+ \right.\\
      &\left. \sum_{k=0}^{\infty} (1-R_{K_{T}k})P_{Q}(q_{i}|k) P(k, \mu_i) \theta(\kappa_i - K_T-1)  \right) \\ 
      &\prod_{T-\textrm{valid}~hit} \left( \frac{\sum^{K}_{k=1} P_{T}(t_{i,r}|r, d_i, \mu^{l}_i, \mu^{d}_i, k) P(k, \mu^{l}_i+\mu^{d}_i)} {\sum^{k}_{k=1} P(k, \mu^{l}_i+\mu^{d}_i)}\right),
\end{split}
\end{equation}

This approach is similar to QTMLE described in Ref.~\cite{QTMLE}. The only difference is that the charge-based likelihood function in Eq.~\ref{eq:likelihood_q} is replaced with the probability vector \{$p^i_k$\} and charge $q_i$ based likelihood function in Eq.~\ref{eq:likelihood_mix}.
We designate this method as Charge, Probability, and Time Maximum Likelihood Estimation (QPTMLE).
The optimal $K_T$ value will be determined by the photon-counting performance and the scanning results discussed subsequently. 

\subsection{Energy reconstruction performances}
\label{sec:Eres}
To assess the potential improvement of the event reconstruction after incorporating the photon counting information of PMTs, 
we use the QTMLE algorithm, as described in Ref.\cite{QTMLE}, as the benchmark.
Since we are mainly interested in the energy in this paper, this section will focus solely on the  energy reconstruction results.
The performance evaluation follows the methodology established in Ref.\cite{QTMLE}, examining energy resolution, non-linearity, and non-uniformity. Detailed definitions of these performance metrics are available in Ref.~\cite{QTMLE}.

For the ideal case where the PMTs have no charge smearing, the true nPEs within the signal window can be obtained. By substituting Eq.\ref{eq:likelihood_q} with Eq.\ref{eq:likelihood_k} in the QTMLE likelihood function, we define a reconstruction algorithm for this ideal case, termed PE and Time Maximum Likelihood Estimation (PETMLE).
This ideal situation is equivalent to 100\% accuracy in machine-learning-based photon counting, a condition challenging to achieve due to overlapping photons and the limited charge resolution of PMTs. However, analyzing the reconstruction performance in this idealized context offers valuable insights.

In the realistic case, the machine-learning-based photon counting model has limited accuracy.
After incorporating the photon counting information into the likelihood function, the reconstruction performance would be expected to fall between the reference and ideal cases.
As mentioned previously, there is another free parameter $K_T$ in the likelihood function Eq.~\ref{eq:EqQPTMLE} for QPTMLE. Its value is scanned starting from 1 and incrementally increased by 1. 

\begin{figure}[htbp]
    \centering
    \includegraphics[width=0.45\textwidth]{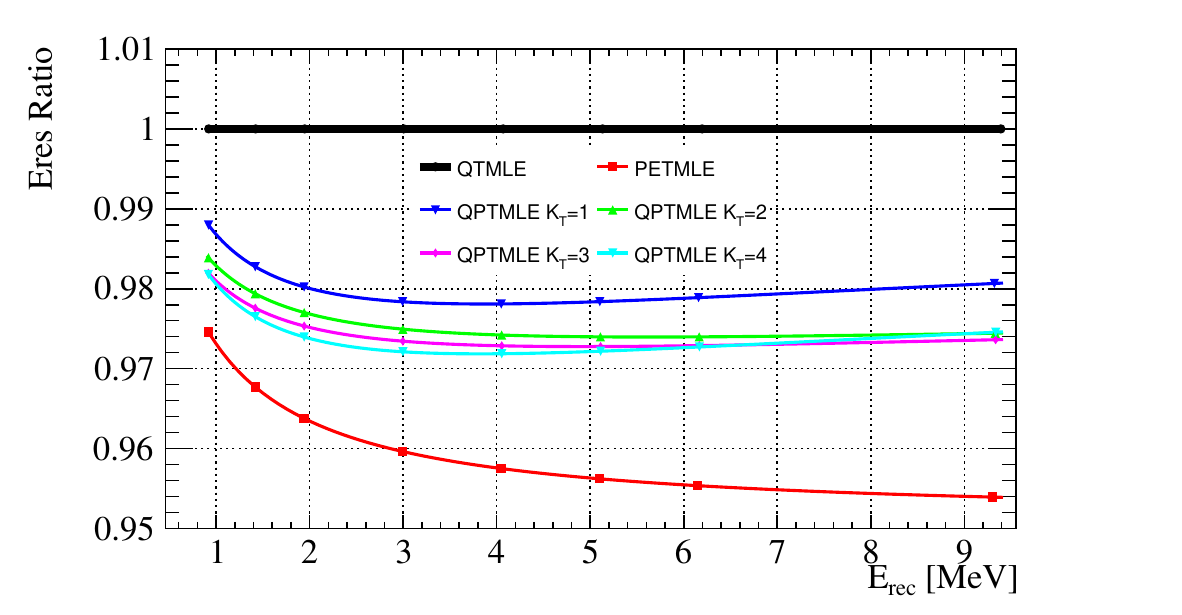}
    \includegraphics[width=0.45\textwidth]{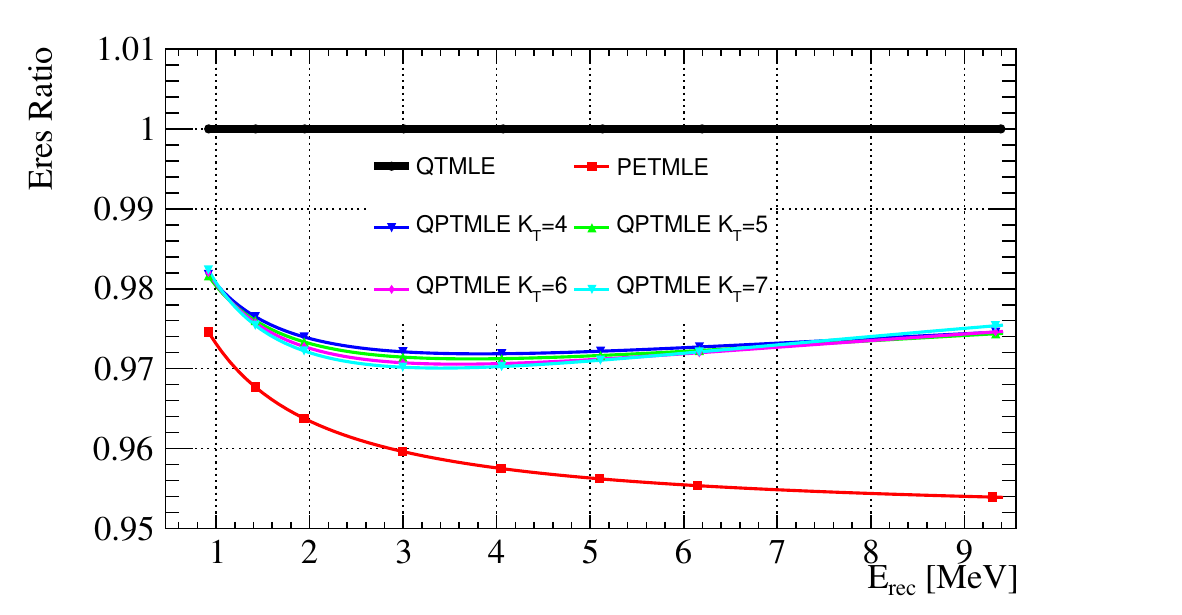}    

    \caption{The comparison of the fitted energy resolution among QTMLE, PETMLE and QPTMLE with different $K_T$. The results of QTMLE are set as the benchmark and the ratio to it from all algorithms are presented. It was observed that the energy resolutions of applying $K_T=4\sim7$ were comparable. QPTMLE-4 demonstrates an improvement in energy resolution by approximately 2.0\%$\sim$2.8\% compared to QTMLE, while the differences between QPTMLE-4 and the ideal case PETMLE are within the range of 0.8\%$\sim$2\%.}
    \label{fig:QPTMLE_res_ratio}
\end{figure}

Table~\ref{tab:comparison_cases} shows the comparison of the three different cases.
\begin{table}[!h]
\centering
\caption{Comparison of the three cases.}
    \begin{tabular}{lccc}
\toprule
       Case & Name     & \textbf{observable} & \textbf{likelihood} \\
\midrule
   reference & QTMLE  &  q &  Eq.~\ref{eq:likelihood_q}  \\
   ideal &  PETMLE &  k & Eq.~\ref{eq:likelihood_k}\\
   realistic & QPTMLE &  $\{p_k\}$, q & Eq.~\ref{eq:likelihood_mix}  \\

\bottomrule
\end{tabular}
\label{tab:comparison_cases}
\end{table}
The comparison of the fitted energy resolution using the positron samples with discrete energies among QTMLE, PETMLE, and QPTMLE with different $K_T$ is illustrated in Fig.~\ref{fig:QPTMLE_res_ratio}.
The black and red curves represent the reference case (QTMLE) and the ideal case (PETMLE), respectively. The realistic case of QPTMLE with different $K_T$ values are also displayed. As mentioned before, the comparison between the ideal and reference cases clearly shows the significant impact of PMT charge smearing on energy resolution.

In the realistic scenario of QPTMLE with $K_T = 1$, the energy resolution significantly improves compared to the reference QTMLE case.
This improvement is attributed to the high proportion of PMTs with a true count of 1~p.e. per event and the 99\% prediction accuracy obtained with the machine learning method for these PMTs.
As $K_T$ increases, the prediction accuracy for PMTs with truth $K_T$~p.e. decreases rapidly, and the contamination due to the PMTs being wrongly predicted as $\kappa=K_T$ grows progressively larger. 
These factors lead to a continuous yet small improvement for the final energy resolution, particularly when compared to the realistic case with $K_T = 1$. 

\begin{figure}[htbp]
    \centering
    \includegraphics[width=0.45\textwidth]{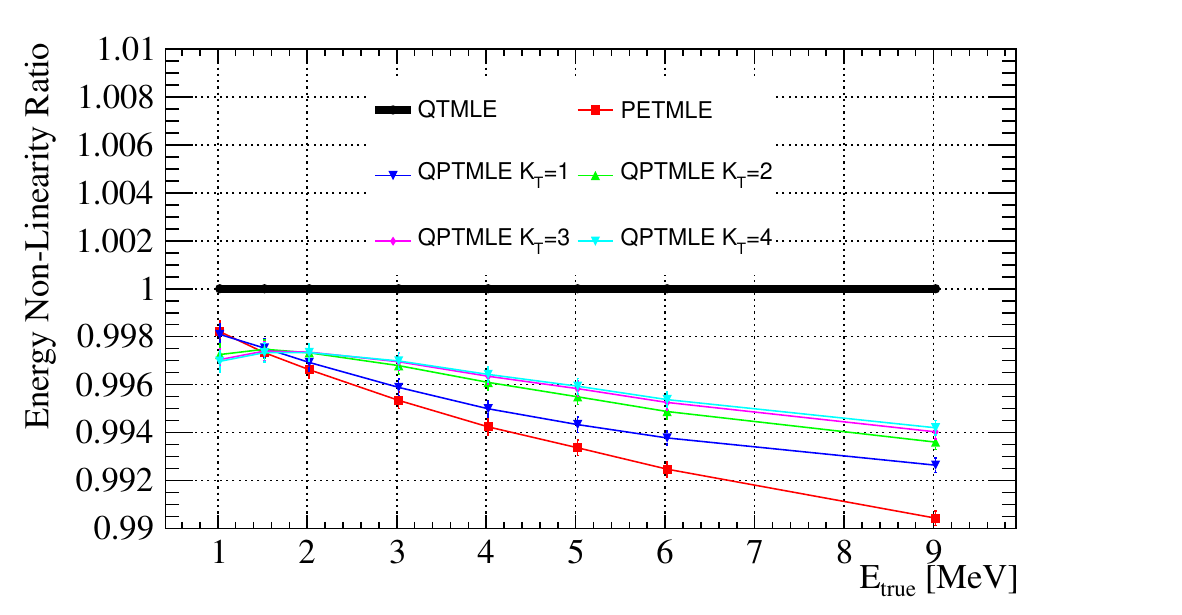}
    \includegraphics[width=0.45\textwidth]{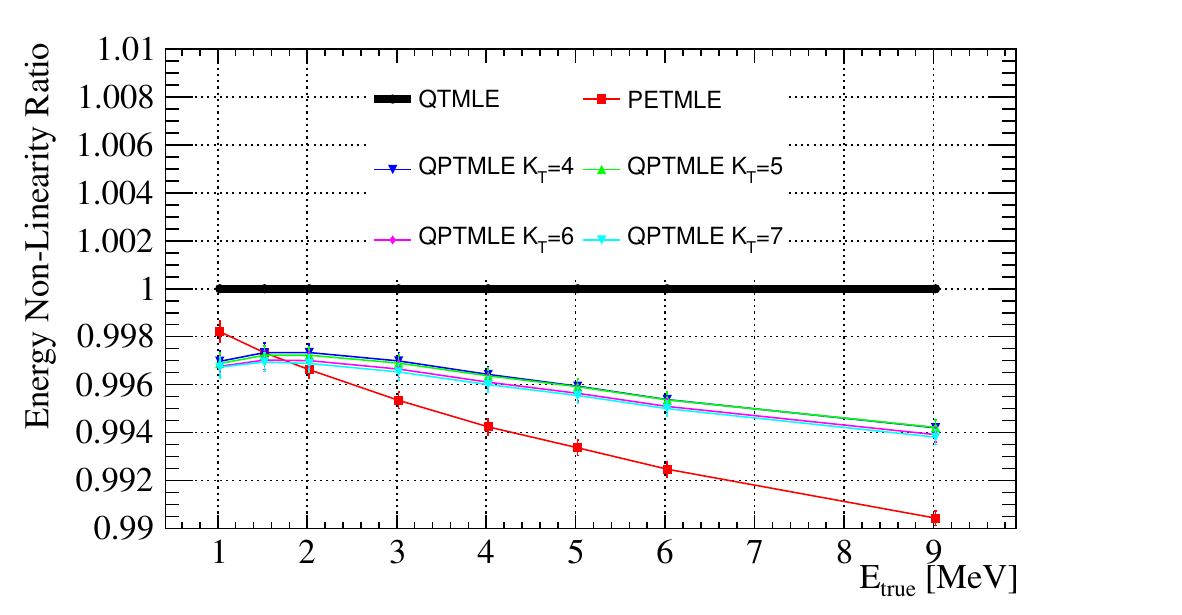}
    \caption{The comparison of energy non-linearity among QTMLE, PETMEL, and QPTMLE. The results of QTMLE are set as the benchmark and the ratio to it from all algorithms are presented. The energy non-linearity of QPTMLE is positioned between QTMLE and PETMLE, The differences of the charge non-linearity among QPTMLE-$K_T$ with $K_T \geq 3$ are marginal.}
    
    \label{fig:comp_nonlinearity}
\end{figure}

\begin{figure}[h]
    \centering
    \includegraphics[width=0.45\textwidth]{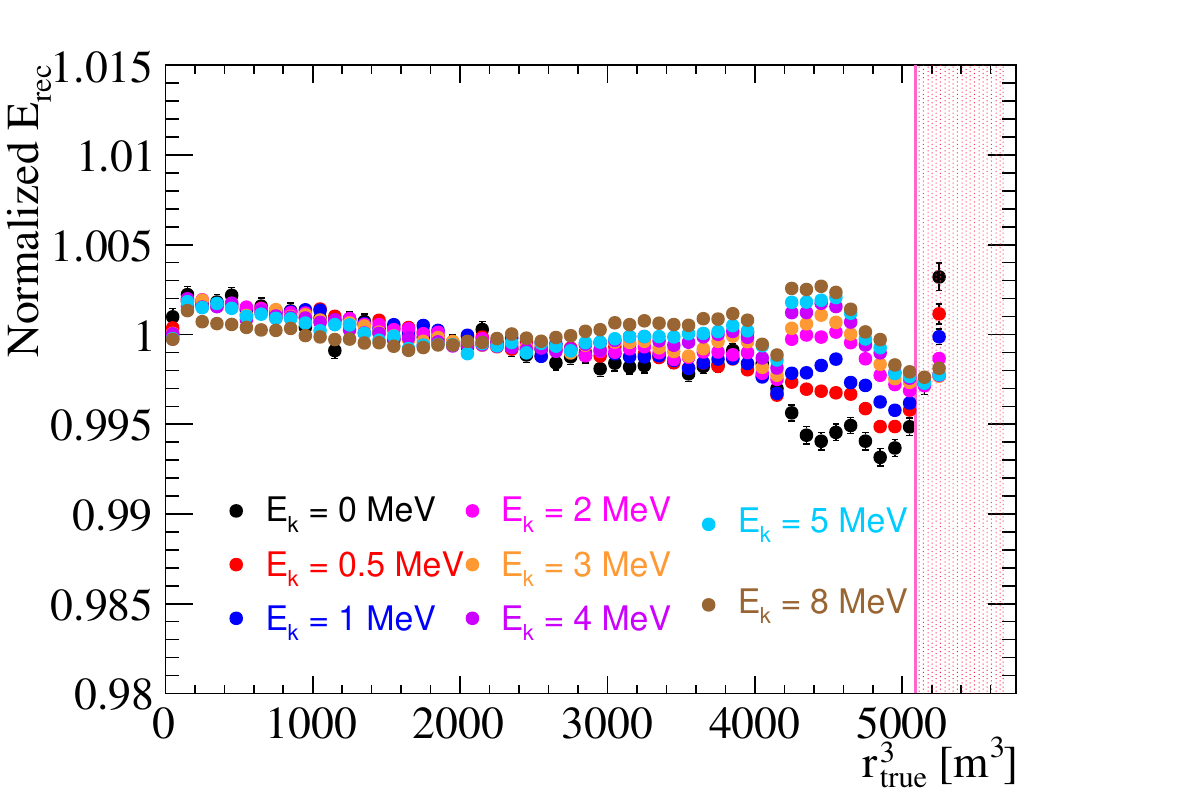}
    \caption{Energy non-uniformity of QPTMLE-4 algorithm. The reconstructed energy is normalized by dividing the average value within the FV. The variation of the average $E_{rec}$ at different positions $r^{3}_{true}$ ranges from 0.3\% to -0.7\% for positron samples of varying energies, which is comparable to that of QTMLE~\cite{QTMLE}}.
    \label{fig:non_uniformity}
\end{figure}

Figure~\ref{fig:comp_nonlinearity} shows the energy non-linearity performance for the reference case QTMLE, the realistic cases QPTMLE-$K_T$, and the ideal case PETMLE. Same as Fig.~\ref{fig:QPTMLE_res_ratio}, the black and red curves correspond to the reference case QTMLE and the ideal case PETMLE, respectively.
By examining both the energy resolution in Fig.~\ref{fig:QPTMLE_res_ratio} and the energy non-linearity in Fig.~\ref{fig:comp_nonlinearity}, it becomes apparent that when $K_T$ exceeds 4, 
the improvement on the energy reconstruction performance becomes marginal. 
Moreover, there is a slight decline of the energy resolution at higher energies for larger $K_T$.
Thus we set the benchmark value for $K_T$ in QPTMLE to 4.

The energy non-uniformity for the realistic case, QPTMLE-4 with $K_T=4$, is shown in Fig.~\ref{fig:non_uniformity}. 
For positrons with different energies, the energy non-uniformity ranges from 0.3\% to -0.7\% within the FV,
comparable to the reference case QTMLE. This indicates that incorporating photon counting information does not adversely affect energy non-uniformity.

\begin{figure}[h]
    \centering
    \includegraphics[width=0.45\textwidth]{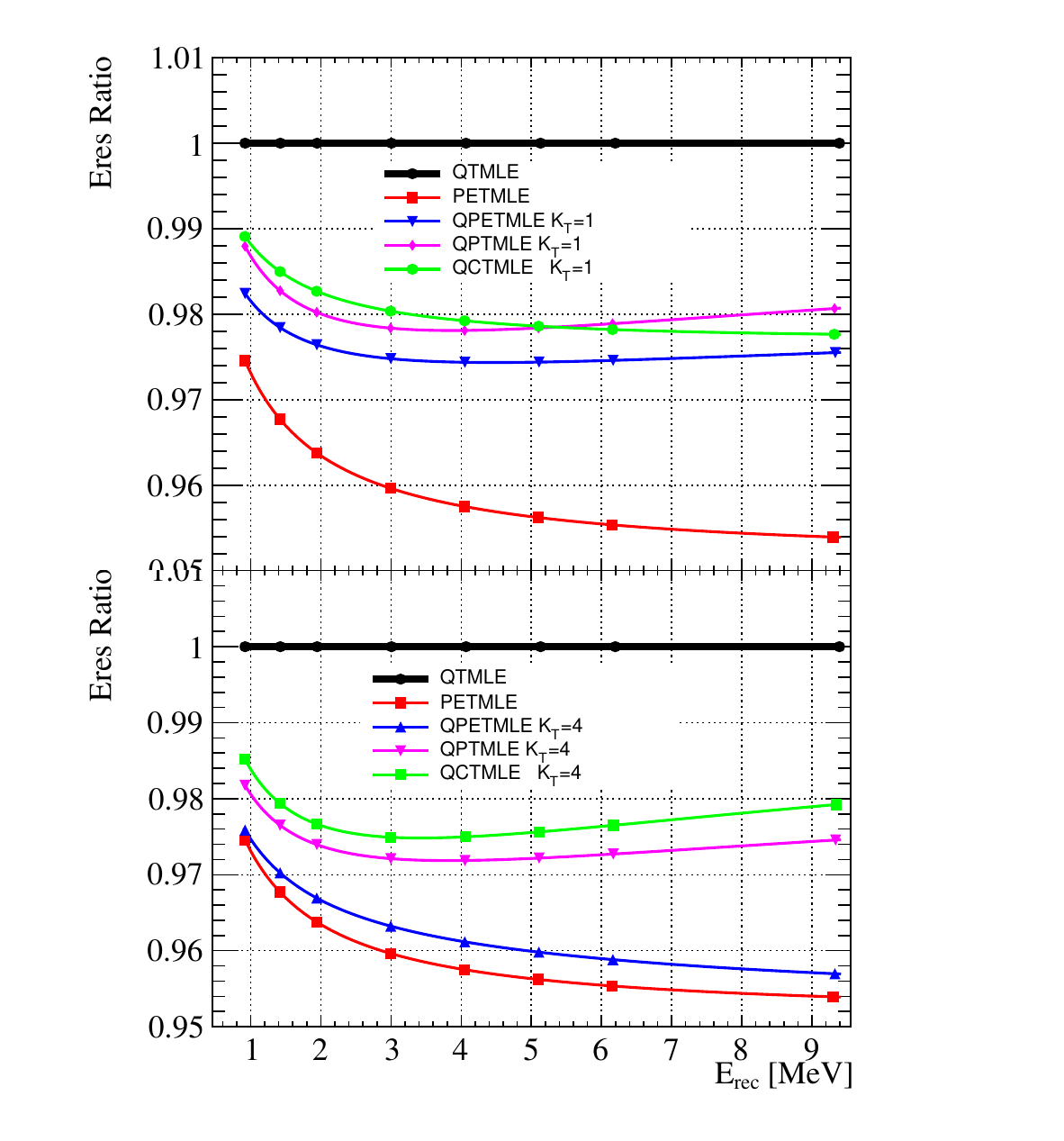}
    \caption{Comparison of the fitted energy resolution of QTMLE, PETMEL, QPETMLE, QPTMLE and QCTMLE (described in Sec.~\ref{sec:discussion} later) with $K_{T}$= 1 and 4. For $K_{T}=1$, QPTMLE has slightly worse energy resolution than QPETMLE. In the case of $K_{T}=4$, the energy resolution of QPTMLE is about 0.6\%$\sim$1.7\% worse than that of QPETMLE from 1 to 8 MeV. QCTMLE exhibits a comparable performance to QPTMLE, with QPTMLE giving slightly better results.}
    \label{fig:PEvsQK_res_ratio}
\end{figure}

\begin{figure}[h]
    \centering
    \includegraphics[width=0.45\textwidth]{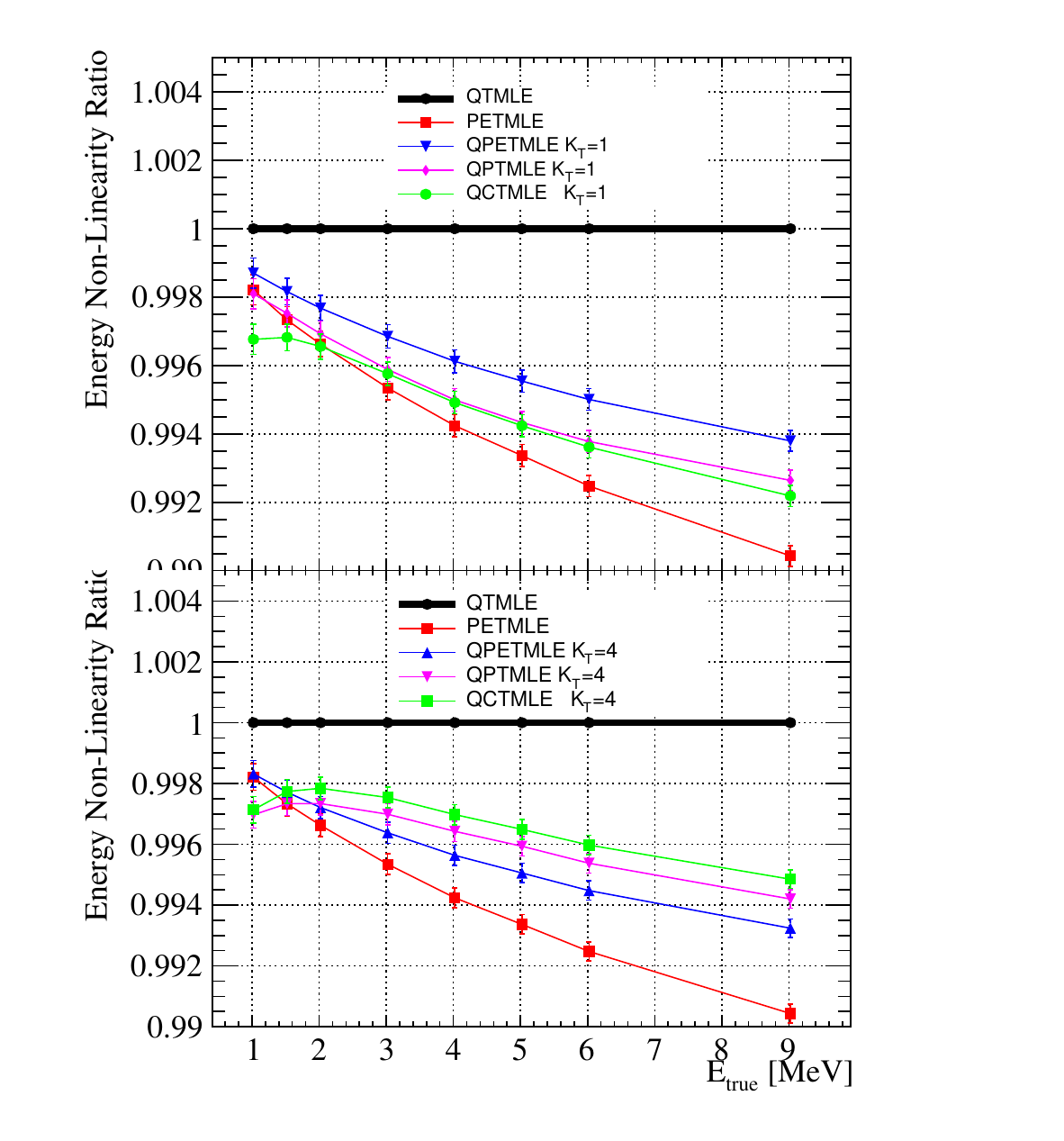}
    \caption{The comparison of energy non-linearity of QTMLE, PETMEL, QPETMLE, QPTMLE and QCTMLE (described in Sec.~\ref{sec:discussion} later). The energy non-linearity of QPTMLE and QCTMLE are comparable to that of QPETMLE. The difference is within 0.2\%. 
    }
    \label{fig:comp4_nonlinearity}
\end{figure}

In order to validate the QPTMLE results, a further check is done using the Charge, PE, and Time Maximum likelihood Estimation, denoted as the QPETMLE-$K_T$ algorithm. This algorithm is similar to QPTMLE-$K_T$, but uses likelihood function Eq.~\ref{eq:likelihood_k} instead of Eq.~\ref{eq:likelihood_p} for PMTs with $\kappa \leq K_T$.
For $K_T$=0, QPETMLE-0 is equivalent to QTMLE, while for $K_T=\infty$, QPETMLE-$\infty$ is equivalent to PETMLE.
The fitted energy resolution of QPETMLE-1 and QPETMLE-4 are compared to those of QPTMLE-1 and QPTMLE-4, as shown in Fig.~\ref{fig:PEvsQK_res_ratio}. 
Given the 99\% accuracy for PMTs with a true count of 1 p.e. and the 3\% misprediction rate for 2~p.e. as $\kappa=1$, QPTMLE-1 has slightly worse energy resolution than QPETMLE-1, yet both algorithms have similar energy non-linearity, as shown in Fig.~\ref{fig:comp4_nonlinearity}. 
As prediction accuracy decreases and contamination increases for PMTs with more PEs, the effectiveness of ML-based photon counting in reducing charge smearing decreases, leading to an increasing discrepancy in energy resolution between QPTMLE and QPETMLE. 
For example, the energy resolution of QPTMLE-1 is approximately 0.5\% worse than that of QPETMLE-1, while for QPTMLE-4, the discrepancy increases to approximately 0.6\% (1.7\%) at low (high) energies.

\section{Alternative approach}
\label{sec:discussion}

As mentioned previously, there are alternative methods for incorporating the photon counting information provided by the machine learning model.
The predicted $\kappa$ value
can be used as a direct estimation of true nPEs for each PMT. 
The probability $P(\kappa|k$) is simply the element $C_{k\kappa}$ in the confusion matrix of the machine learning model,
representing the probability of a PMT with k p.e. being predicted as $\kappa$ p.e.. 
For PMTs satisfying the $\kappa \leq K_T $ requirement, the likelihood function is constructed using Eq.~\ref{eq:likelihood_c}:

\begin{equation}
\label{eq:likelihood_c}
    \mathcal{L}( \kappa_{i} | \mu_i) = 
    \sum_{k} P(\kappa_i|k) \times P(k, \mu_i) =
    \sum_{k=0}^{9} C_{k\kappa_i} \times P(k, \mu_i) 
\end{equation}

\noindent which is very similar to Eq.~\ref{eq:likelihood_p}. Again for $C_{k9}$, $P(9, \mu_i)$ is defined as $\sum_{k=9}^{+\infty} P(k, \mu_i)$. The crucial difference lies in the utilization of the overall PMT nPEs response, represented by the confusion matrix from the machine learning model, rather than the individual \{$p_k$\} information for each PMT in the positron samples. 
For PMTs with $\kappa > K_T $, the likelihood function remains the same as QPMLE. And this method is denoted as Charge, Confusion-Matrix and Time Maximum likelihood Estimation (QCTMLE).

Table~\ref{tab:comparison_3} shows a comparison of the strategies discussed. A common aspect between them is the treatment of PMTs with $\kappa > K_T $ and the usage of $\kappa$ to determine the appropriate likelihood function.
\begin{table}[h]
\centering
\caption{Comparison of the three strategies.}
    \begin{tabular}{lccc}
\toprule
       Strategy     & \textbf{ photon counting} & \textbf{likelihood} & \textbf{likelihood}\\
            & \textbf{information used} & $\kappa \leq K_T$ & $\kappa > K_T$\\
\midrule
    QPTMLE  & $\{p_k\}$ &  Eq.~\ref{eq:likelihood_p} &  Eq.~\ref{eq:likelihood_q} \\
    QCTMLE & $C_{k,\kappa}$ & Eq.~\ref{eq:likelihood_c} & Eq.~\ref{eq:likelihood_q} \\

\bottomrule
\end{tabular}
\label{tab:comparison_3}
\end{table}
For the QPTMLE strategy, its likelihood function is for PMTs with $\kappa \leq K_T $ essentially a superposition of the one in Eq.~\ref{eq:likelihood_k} with the weights \{$p_k$\} provided by the machine learning model. The QCTMLE strategy resembles QPTMLE but substitutes the per-PMT \{$p_k$\} information with the overall confusion matrix $C_{k\kappa}$ from the machine learning model.

The comparison of energy resolution and non-linearity obtained by the two strategies are shown in Fig.~\ref{fig:PEvsQK_res_ratio} and Fig.~\ref{fig:comp4_nonlinearity}, respectively. Overall, the performances are comparable, with QPTMLE giving slightly better results. QPTMLE holds an advantage over QCTMLE, as it 
uses the \{$p_k$\} information on an individual PMT basis, effectively capturing the unique nPEs response of each PMT. In contrast, QCTMLE does not fully account for these individual PMT differences. 
For example, for PMTs with the same $\kappa$, the same $C_{k\kappa}$ will be used in QCTMLE, but \{$p_k$\} in QPTMLE will be different.

\section{Conclusion}
\label{sec:summary}
One of the main challenges in determining the Neutrino Mass Ordering (NMO) for JUNO is achieving the required unprecedented energy resolution. The charge smearing effect of PMTs emerges as a dominant factor affecting this resolution. In this paper, a novel machine-learning-based method is developed to predict the nPEs in PMT waveforms. We integrate this model into the event reconstruction process for positrons, combining the photon counting information from PMTs with low PEs into the maximum likelihood estimation for positron energy. 
Different strategies for utilizing the photon counting information are explored. 
The QPTMLE-4 algorithm, which combines the (\{$p_k$\}, k) information for PMTs with $K_T \leq 4$, is set as the benchmark method.
In JUNO, where reactor anti-neutrinos have energies below 10~MeV, most of the PMTs detect only a few photons. Meanwhile the prediction accuracy of the machine-learning-based photon counting model is very high for PMTs with low PEs. The usage of the photon counting information provided by the machine learning model was able to partially mitigate the charge smearing effect of PMTs, and the strategy QPTMLE-4 improves the energy resolution relatively by about 2.0\% to 2.8\% in the energy range of [1, 9]~MeV. In principle, this machine-learning-based photon counting method and its associated utilization strategies could also be applied to other low-energy experiments with many PMTs.

\section*{Acknowledgements}
This work was partially supported by the National Key R\&D Program of China (2023YFA1606103), 
by the National Natural Science Foundation of China (Grant No.12175257, No.12405231), 
and by the Science Foundation of High-Level Talents of Wuyi University (2021AL027).

\end{document}